\documentclass[aps,notitlepage,pre]{revtex4-1}
\usepackage{ucs}
\usepackage[utf8x]{inputenc}
\usepackage{amsmath}
\usepackage{amsfonts}
\usepackage{amssymb}
\usepackage{amsthm}
\usepackage{fontenc}
\usepackage[dvips,pdftex]{graphicx}
\usepackage{setspace}
\usepackage{color}
\usepackage[dvipsnames]{xcolor}
\usepackage{graphicx}

\usepackage{hyperref}

\usepackage{multirow}
\usepackage{array}
\usepackage{caption}
\usepackage{subcaption}
\usepackage{natbib}
\usepackage{multibib}
\newcites{apx}{es}

\usepackage{tikz}
\usepackage{multimedia}
\usepackage{float}
\usepackage{rotating}
\usepackage[outdir=./]{epstopdf}

\makeatletter
\@addtoreset{chapter}{part}
\makeatother

\newcommand{\dt}{\,\text{d}}

\newcommand{\pd}{\partial}
\newcommand{\Tr}{\text{Tr}}
\newcommand{\nablab}{\bar{\nabla}}
\newcommand{\Bb}[2]{\mathbf{B}\left(#1,#2\right)}
\newcommand{\Bbs}[2]{\mathbf{B}^2\left(#1,#2\right)}

 \DeclareGraphicsExtensions{.pdf,.eps,.jpg}

\hypersetup{linktocpage}

\begin{document}

\begin{abstract}
	We suggest a geometrical mechanism for the ordering of slender filaments inside non-isotropic containers, using cortical microtubules in plant cells and packing of viral genetic material inside capsids as concrete examples. We show analytically how the shape of the cell affects the ordering of phantom, non-self-avoiding, elastic rods. We find that for oblate cells the preferred orientation is along the equator, while for prolate spheroids with an aspect ratio close to one, the orientation is along the principal (long axis). Surprisingly, at high enough aspect ratio, a configurational phase transition occurs, and the rods no longer point along the principal axis, but at an angle to it, due to high curvature at the poles. We discuss some of the possible effects of self avoidance, using energy considerations. These results are relevant to other packing problems as well, such as spooling of filament in the industry or spider silk inside water droplets.
\end{abstract} 
\title{On the Packing of Stiff Rods on Ellipsoids Part I - Geometry}
\author{Doron \surname{Grossman}}
\email[]{doron.grossman@mail.huji.ac.il}
\affiliation{Racah Institute of Physics, Hebrew University, Jeruslaem 9190401, Israel}
\author{Eytan \surname{Katzav}}
\email[]{eytan.katzav@mail.huji.ac.il }
\affiliation{Racah Institute of Physics, Hebrew University, Jeruslaem 9190401, Israel}
\author{Eran \surname{Sharon} }
\email[]{erans@mail.huji.ac.il }
\affiliation {Racah Institute of Physics, Hebrew University, Jeruslaem 9190401, Israel}	
\date{\today}
\maketitle
Packing of filaments inside volumes of a given shape is an important problem in many fields. From packing of genetic material inside viral capsids\cite{marenduzzo2010biopolymer,katzav2006statistical,boue2007folding,purohit2003mechanics,angelescu2008viruses}, through ordering of micro-tubules near plant cell walls\cite{uyttewaal2012mechanical,burk2002alteration,stoop2011morphogenesis}, to spooling of filaments in industrial application \cite{vetter2013finite,vetter2015growth,pineirua2013spooling} and other biopolymers such as spider silk \cite{elettro2015elastocapillary,elettro2017drop,elettro2017elastocapillary,elsner2012spatiotemporal}.  These type of problems are hard as they typically involve a compromise between elastic energy and entropic constraints. As a result, most approaches to these problems involve highly symmetric volumes and are mostly numerical. This paper's aim is to gain further analytical understanding of systems which are not highly symmetric, using differential geometry, study of entropic efects and self avoidance is largely postponed to a later paper. As  concrete examples,  we consider plant cells where shape regulation is important and microtubules play a significant role and the packing of genetic material inside viral capsids. While these two examples seem very different, both are essentially problems of packing long semi-flexible filaments.  Additionally, the authors hope that this work will serve as an inspiration for applying similar methods from differential geometry to study other systems, and not only the packing of filaments on surfaces.

In the  context of plant tissue, the confinement of microtubules (MT) in the cell membrane, and especially, the possible transition from disordered to partially ordered packing is thought to be the origin of important morphogenetic and growth regulating mechanisms. The location of leaves on the stem, anisotropic cell expansion, as in roots and stems, and the evolution of anisotropic mechanical properties of extended leaf tissue are few examples in which mechanical and geometrical asymmetries appear.  In plant cells, cellulose fibrils in the cell wall are the carrier of mechanical load. Orientation of cortical MT dictates the orientation of cellulose fibrils deposition in the cell wall \cite{burk2002alteration}. Therefore, anisotropic MT orientation would lead to anisotropic cellulose deposition, and, thus, to anisotropic mechanical properties of the cell. This, in turn, would lead to its anisotropic growth and to shape evolution of the tissue. But what sets MT orientation at the single cell level? The suggestion that mechanical stress directly affects MT orientation, was supported by pointing to the correlations between MT orientation and the directions of principal stresses in evolving meristems	 \cite{uyttewaal2012mechanical}.
On the other hand, it is not clear how stress or strain in the cell wall can affect the cell membrane \cite{fisher1998extending,somerville2004toward}. In addition, the effect of applied stress on mechanical properties of a leaf appeared only after a delay of several hours {\cite{sahaf2016rheology}}. This suggests that an integral mechanism, which is common in other growth regulation mechanisms {\cite{oh2014cell,chaiwanon2016information}}, could be in action. Alternative approaches, suggest that the cell shape, and particularly, its anisotropy is a dominant factor in MT ordering. The two approaches are, in fact, closely related: Considering a circular cell, an integral over the strain determines the cell's anisotropy. We therefore pose the questions of what are the leading effects of cell shape on the disordered-aligned transition of MT. 

In the context of viral capsids there is ultimate necessity for the genetic material to be orderly packed inside the capsid in order to assure effective infection. This is so important because once the genetic material is packed inside the capsid it has no active sources of energy to support its injection, except for the elastic energy stored in the structure. It has been convincingly argued \cite{marenduzzo2010biopolymer,katzav2006statistical,purohit2003mechanics} that the size of the capsid plays an important role in piloting towards an ordered structure based on both entropic and energetic considerations, showing that active biological control over ordering the spool could be seriously reduced. In particular, a continuous phase transition between a disordered phase (for large capsids) and an ordered phase (for small capsids) has been found.
The spontaneous formation of virus capsids from multiple copies of capsid proteins is itself a fascinating example of supramolecular self-assembling processes \cite{fejer2020minimalistic}. Most known viruses protect their genome with icosahedral capsids, which is often modelled as a spherical cavity. However, other capsid morphologies exist as well, notably elongated, often ellipsoid, structures. The bacteriophage $\phi 29$ has elongated capsids with size of about $42{\rm nm} \times 54{\rm nm}$ leading to an aspect ratio of $4:5$ \cite{marenduzzo2009statistics,ali2006polymer}. The capsid of archaeal virus Sulfolobus ellipsoid virus 1 (SEV1) \cite{wang2018novel} is that of a prolate ellipsoid with principal axes of $78{\rm nm} \times 115{\rm nm}$ hence aspect ratio of about $2:3$. Other important examples include the human virus, HIV-1, where experiments have demonstrated the self-assembly of viral capsid proteins into elongated shapes both in vitro and in vivo \cite{krishna2010role} with varying aspect ratios depending on the precise conditions during the assembly process, notably the interaction between the building block proteins. An interesting example is the case of extremely elongated capsids in mutant T4 phages \cite{rao1993purification} with aspect ratios in the range $1:2$-$1:3$ leading to a twisted toroidal configuration of DNA inside them.
An immediate question that comes to mind is: does the shape of the capsid affect the ordering phase transition of the genetic material in the capsid compared to the spherical case? Simulations performed by different groups \cite{marenduzzo2010biopolymer,vetter2013finite,marenduzzo2008computer,marenduzzo2009statistics} indicate that packings in oblate spheroids and scalene ellipsoids are energetically preferred to spheres. In simple words, it seems that the elongated container tends to order the structure more than a spherical one. Furthermore, it was observed in simulations \cite{marenduzzo2009statistics,ali2006polymer,petrov2007conformation,petrov2008packaging,petrov2009characterization} that the kind of order of the genetic material changes, and attains a nematically ordered structure with twist (also called twisted toroidal conformations \cite{petrov2007conformation}), as opposed to a spool-like one in spherical capsids. The different simulations are not clear about the source of this effect. Some suggest that the toroidal packing arranges itself along the longest possible dimension to minimize the bending stress (and hence the elastic energy) \cite{petrov2007conformation}, while others put more emphasis on entropic effect \cite{jun2006entropy}.
In this paper we want to clarify the energetic aspects of packing an elastic rod, and in particular the enhanced ordering emanating from the elongated geometry, while in a future publication we will focus on entropic effects.
  
 We begin by assuming a a phantom ,i.e. non - self - avoiding, unstretchable, filament of thickness $t$ whose
 midline is given by ${\bf R}(s) = \left\{X(s),Y(s),Z(s)\right\}$ where $s$ is the arclength along the rod. The bending energy of such a rod is 
\begin{align}\label{eq:rod_ener}
 E & = \frac{{\varepsilon} t^4}{2} \int  \left| \ddot{\bf R}(s)\right|^2 \dt s = \frac{{\varepsilon} t^4}{2} \int \left( \ddot{X}^2 +\ddot{Y}^2 +\ddot{Z}^2\right) \dt s
\end{align}
 where $\ddot{Q} = \pd_s^2 {Q}$ is the second derivative of the quantity $Q(s)$, and ${\varepsilon}$ is the rod's Young's modulus. At zero temperature, the rod's configuration is the global minimizer of Eq.\eqref{eq:rod_ener}, subjected to the constraint of lying on the surface of an ellipsoid - $X^2(s) +Y^2(s)+ Z^2(s)/(1+\delta)^2 =1$, where $\delta> -1$ is called the flattening of the ellipsoid, and is related to the usual definition of eccentricity $e$, via $e^2 = \delta(2+\delta)$ \cite{newton1726philosophiae,bessel2009uber,bessel2009calculation}. A vanishing flattening, $\delta=0$, corresponds to a sphere. Negative values of $\delta$ in the range $-1<\delta<0$ describe an oblate spheroid, while positive values, $\delta>0$ is a prolate spheroid. The problem of finding the global energy minimizer is difficult, as the system is clearly highly frustrated. We therefore use a geometrical approach, and describe the rod using its curvatures rather than the configuration $\mathbf{R}$. We start by assigning a generalized Frenet - Serret frame (see \cite{Panyukov2000,Grossman2016,Grossman2018shape}), $\left\{ \mathbf{t}_1, \mathbf{t}_2, \mathbf{t}_3\right\}$, being the tangent, bi-normal and normal vectors of the rod, respectively, as follows.
Given $\mathbf{f}= \left\{X(x,y),Y(x,y),Z(x,y)\right\}$, the surface configuration ($x,y$ being coordinates \emph{on the surface}), and $\mathbf{\hat{n}}$ the normal to the surface, we choose a  local orthonormal in-plane frame along the rod, using a Darboux frame on the surface $\left\{ \pd_x \mathbf{f}, \pd_y \mathbf{f}, \mathbf{\hat{n}}\right\}$ we define the Frenet - Serret frame, $\left\{ \mathbf{t}_1, \mathbf{t}_2, \mathbf{t}_3\right\}$, and the in - plane surface tangent and normal vectors $e_1^\mu$ and $e_2^\mu$:
\begin{subequations}	
	\begin{align}
	\mathbf{t}_1 &=\dot{\bf R} = e_1^\mu \pd_\mu \mathbf{f}\\
	\mathbf{t}_2 &= e_2^\mu \pd_\mu \mathbf{f}\\
	\mathbf{t}_3 &=  \mathbf{\hat{n}}.
	\end{align}
	 Orthonormality dictates, that the inner products satisfy $e_1^\mu e_{1 \mu}=a_{\mu\nu}e_1^\mu e_1^\nu=1= e_2^\mu e_{2 \mu}$, and $e_1^\mu e_{2\mu}=0$, where $a_{\mu \nu}= \pd_\mu \mathbf{f} \cdot \pd_\nu \mathbf{f}$ is the induced metric on the surface, and $\mu, \nu \in \left\{x,y\right\}$.
\end{subequations}
It is then straightforward to show that $\mathbf{t}_{\alpha}$, $\alpha \in \{1,2\}$ satisfy the equations,
\begin{align}\label{eq:gen_Fre_Ser1}
\dot{\mathbf{t}}_\alpha &=\partial_s \mathbf{t}_\alpha=  (\pd_s e_\alpha ^\mu )\pd_\mu \mathbf{f} + e_1^\mu e_\alpha^\nu \partial_\mu\partial_\nu \mathbf{f} =  -\kappa_g \sum_\beta \epsilon_{\alpha \beta}e_\beta^\mu \partial_\mu\mathbf{f} + e_1^\mu e_\alpha^\nu b_{\mu\nu} \mathbf{t}_3,
\end{align}
and $\mathbf{t}_3$ satisfies
\begin{align} \label{eq:weingarten}
\dot{\mathbf{t}}_3 = - \sum_{\alpha} e_1^\mu e_\alpha^\nu b_{\mu \nu} \mathbf{t}_{\alpha}.
\end{align}
Here $b_{\mu \nu}=  \pd_\mu \pd_\nu \mathbf{f} \cdot \hat{\mathbf{n}}$ is the second fundamental form of the confining surface, $\kappa_g$ is the geodesic curvature of the rod on the surface, and $\epsilon_{\alpha \beta}$ is the $2\times2$ antisymmetric (Levi-Civita) symbol. Deriving Eq. \eqref{eq:gen_Fre_Ser1}, we used the fact that for a general curve, namely a non geodesic one, on the surface $e_1^\mu$ and $e_2^\mu$ are transported via the equation 
\begin{align}\label{eq:non_geo_eq}
\nabla_s e_\alpha^\mu &=\pd_s e_\alpha^\mu + e_1^\rho e_\alpha^\sigma \Gamma_{\rho \sigma}^\mu =  -\kappa_g \sum_\beta \epsilon_{\alpha \beta}\, e_\beta^\mu
\end{align}
where $\nabla_s$ is the covariant derivative along the curve, and the Christoffel symbols, given a configuration 
\begin{align}\label{eq:def_christoffels}
a_{\rho \lambda} \Gamma_{\mu\nu}^\lambda = \pd_\mu\pd_\nu \mathbf{f} \cdot \pd_\rho\mathbf{f}.
\end{align}
Eqs. \eqref{eq:gen_Fre_Ser1} and \eqref{eq:weingarten} can be rewritten in matrix form as
\begin{align}\label{eq:gen_Fre_Ser2}
\left(\begin{array}{c}
\dot{\mathbf{t}}_1(s) \\ \dot{\mathbf{t}}_2 (s)\\\dot{\mathbf{t}}_3(s)
\end{array}\right) = -\left( \begin{array}{c c c}
0 & \kappa_g  & -l\\
-\kappa_g & 0 & m \\
l &-m & 0
\end{array}\right) \left(\begin{array}{c}
\mathbf{t}_1(s) \\ \mathbf{t}_2 (s)\\\mathbf{t}_3(s)
\end{array}\right)
\end{align}
where we defined the curvatures
\begin{subequations}
	\begin{align}
	l &= e_1^\mu e_1^\nu b_{\mu\nu} \\
	m &= -e_1^\mu e_2^\nu b_{\mu\nu}
	\end{align}
\end{subequations}
The formal solution of Eq. \eqref{eq:gen_Fre_Ser2} is
\begin{align}  \left(\begin{array}{c}
\mathbf{t}_1(s) \\ \mathbf{t}_2 (s)\\\mathbf{t}_3(s)
\end{array}\right) = T_s\left[ \exp\left(- \int\limits_0^s \dt s'\left( \begin{array}{c c c}
0 & \kappa_g  & -l\\
-\kappa_g & 0 &m \\
l &-m & 0
\end{array}\right) \right)\right] \left(\begin{array}{c}
\mathbf{t}_1(0) \\ \mathbf{t}_2 (0)\\\mathbf{t}_3(0)
\end{array}\right)
\end{align}
where we use the definition of the position ordering operator $T_s$
 \begin{align}
 T_s \exp\left[{ \int\limits_{0}^s M \dt s'}\right] = \lim\limits_{k\rightarrow \infty} e^{M(s_0) \Delta s} e^{M(s_1) \Delta s}\dots e^{M(s_{k}) \Delta s},
 \end{align}  where $\Delta s = \frac{s}{k}$, and $s_k = k \Delta s$.
The rod configuration is then given by
\begin{align}
 \mathbf{R}(s)= \mathbf{R}(0)+ \int\limits_0^s \mathbf{t}_1(s') \dt s'
\end{align}
The elastic energy \eqref{eq:rod_ener} assumes the form
\begin{align}\label{eq:rod_ener_int}
E & = \frac{\varepsilon t^4}{2} \int \dt s \left( \kappa_g^2 + l^2 	 \right).
\end{align}
While this seems to be a simpler functional than the one given in Eq. \eqref{eq:rod_ener}, one must remember that $l$ itself is a functional of the geodesic curvature, $\kappa_g(s)$ (via Eq. \eqref{eq:non_geo_eq}).
Since we are searching for global minimizers (for infinitely long rods) we seek to minimize the total energy per unit length.
\begin{align}\label{eq:ener_glob}
\bar{E} &=\frac{E}{L} = E_0\frac{\displaystyle\int \left(\kappa_g^2  + l^2\right) \dt s}{\displaystyle\int \dt s},
\end{align}
{
where we define the energy scale $E_0 =\frac{\varepsilon t^4}{2}$.

 Eq. \eqref{eq:ener_glob} reveals an important property of the energy minimizing problem: while the normal curvature, $l$, is a property of the surface, the geodesic curvature, $k_g$, can be eliminated by a proper selection of the rod configuration. These configurations are geodesics, which locally minimize Eq. \eqref{eq:ener_glob}. As shown later (appendix \ref{appndx: stability}), for any positive flattening, $\delta>0$, no geodesic is a global minimizer. However, the true global minimizer cannot deviate significantly from  a geodesic (see appendix \ref{appndx: deviation estimation}). Therefore, we conclude geodesics are good approximations,an limit ourselves to their study, which significantly simplifies the problem.

}

We begin by parametrising the surface of the ellipsoid using $\theta$ (polar) and $\phi$ (azimuthal) angles, and subsequently finding the geodesics in terms of these coordinates:
  \begin{align} \label{eq:parametrization} \nonumber
  X&=\cos\phi\sin\theta\\
  Y&=\sin\phi\sin\theta\\ \nonumber
  Z&=(1+\delta)\cos\theta
  \end{align}
  The metric $\mathbf{a}$, the second fundamental form $\mathbf{b}$ and the shape operator (extrinsic curvature) $\mathbf{S}$ \cite{do2016differential} of the surface are then given by

  \begin{align} \label{eq:metric_shape}
  \mathbf{a}&= \left(\begin{array}{cc}
  1+\delta (2+\delta)\sin^2\theta & 0 \\
  0 & \sin^2\theta
  \end{array}\right) 
\\
  \mathbf{b}&= -\frac{1+\delta}{\sqrt{1+\delta(2+\delta)\sin^2\theta}} \left(\begin{array}{cc}
  1 & 0 \\
  0 & \sin^2\theta
  \end{array}\right) 
\\
\mathbf{S}&=\mathbf{a}^{-1}\mathbf{b} =-\left(\begin{array}{cc}
\frac{1+\delta}{({1+\delta(2+\delta)\sin^2\theta})^{3/2}} & 0 \\
0 & \frac{1+\delta}{\sqrt{1+\delta(2+\delta)\sin^2\theta}}
\end{array}\right).
  \end{align}   
 
Since the metric is independent of $\phi$, there is a  conserved, conjugate ``angular momentum" $\omega$, which is related to the ``angular velocity" $\dot{\phi}$ , and is the generator of rotations around the symmetry axis of the ellipsoid found by solving the geodesic equation for $\phi$ coordinate \cite{do2016differential}:
\begin{align}
\frac{d}{ds}\left(a_{\phi \mu} \frac{d x^\mu}{ds}\right) &= 0  \\ 
\intertext{Since the metric is diagonal, only $a_{\phi\phi}= \sin^2\theta$ contributes to the sum -}
\frac{d}{ds}\left(a_{\phi \phi} \frac{d \phi}{ds}\right) &= 0 \\ 
\intertext{in other words, this is a constant, which we mark by $\omega$}
\sin^2\theta \frac{d \phi^\mu}{ds} &= \omega =const.
\end{align}
Thus
\begin{align} \label{eq:phi_geo}
 	\dot{\phi}(s)&=\frac{\omega}{\sin^2\theta},
\end{align}
where arc-length normalization of $s$ requires $-1\leq\omega\leq1$.
Instead of solving the geodesic equation for $\theta$ we may impose normalization, $e^\mu_1 e_1^\nu a_{\mu \nu}=1$ and get  the equation
\begin{align}
	\label{eq:thet_geo_dif}
	\left(1+ 2\delta \sin^2\theta +\delta^2 \sin^2\theta \right)  \dot{\theta}^2 &= 1-\frac{\omega^2}{\sin^2\theta}.
	\end{align}
	By taking $\sin^2 \theta$ as a common factor we get
	\begin{align}
	\left(\frac{1}{\sin^2\theta}+ 2\delta  +\delta^2  \right) \sin^2\theta \dot{\theta}^2 &= 1-\frac{\omega^2}{\sin^2\theta}
	\end{align}
	leading to
	\begin{align}
	\left(\frac{d}{ds} \cos\theta\right)^2 &= \frac{1-\frac{\omega^2}{\sin^2\theta}}{\left(\frac{1}{\sin^2\theta}+ 2\delta  +\delta^2  \right)}.
	\end{align}
	Rewriting everything in terms of $\cos\theta$ we get the equation
	\begin{align}
	\left(\frac{d}{ds} \cos\theta\right)^2 &= \frac{1-\omega^2-\cos^2\theta}{\left[(1+\delta)^2- 2\delta \cos^2\theta  -\delta^2  \cos^2\theta \right]}.
	\end{align}
	Which after defining $x(s)= \cos\left[\theta(s)\right]$, and marking $\dot{x} = \frac{d x}{d s}$ turns into
	\begin{align}
	(\dot{x})^2 &= \frac{1-\omega^2-x^2}{(1+\delta)^2-( 2\delta+\delta^2) x^2}.
	\end{align}
	Separating variables and integrating we obtain
	\begin{align}
	\int \frac{\sqrt{(1+\delta)^2-( 2\delta+\delta^2) x^2}\dt x}{\sqrt{1-\omega^2-x^2}} &= \pm \int \dt s + s_0.
	\end{align}
	This equation is integrable, giving rise to
	\begin{align}
 \label{eq:thet_geo}
	(1+\delta) \mathcal{E}\left[\sin ^{-1}\left(\frac{x}{\sqrt{1-\omega ^2}}\right)\Bigg|\frac{\delta  (\delta +2) \left(1-\omega ^2\right)}{(\delta +1)^2}\right] &= \pm s+s_0
\end{align}
where $\mathcal{E}[\theta|m]= \int\limits_{0}^\theta \sqrt{1-m \sin^2\phi} \dt \phi$ is the incomplete elliptic integral of the second kind \cite{olver2010nist}, and $s_0$ is determined by initial conditions. To find geodesics we now need to integrate Eq. \eqref{eq:phi_geo}, which can be done numerically. However, by virtue of rotational symmetry about the polar axis of the ellipsoid ($a_{\mu\nu}$ is independent of $\phi$, geodesics are parametrized by $\omega$ alone), we can disregard exact solutions of $\phi(s)$ when looking for minimizing geodesics, and we can immediately look for a minimizer of Eq. \eqref{eq:ener_glob} (see Eq. \eqref{eq:ener_glob_geo}).

\begin{figure}[!h]
\centering	
\includegraphics[width=0.7\textwidth]{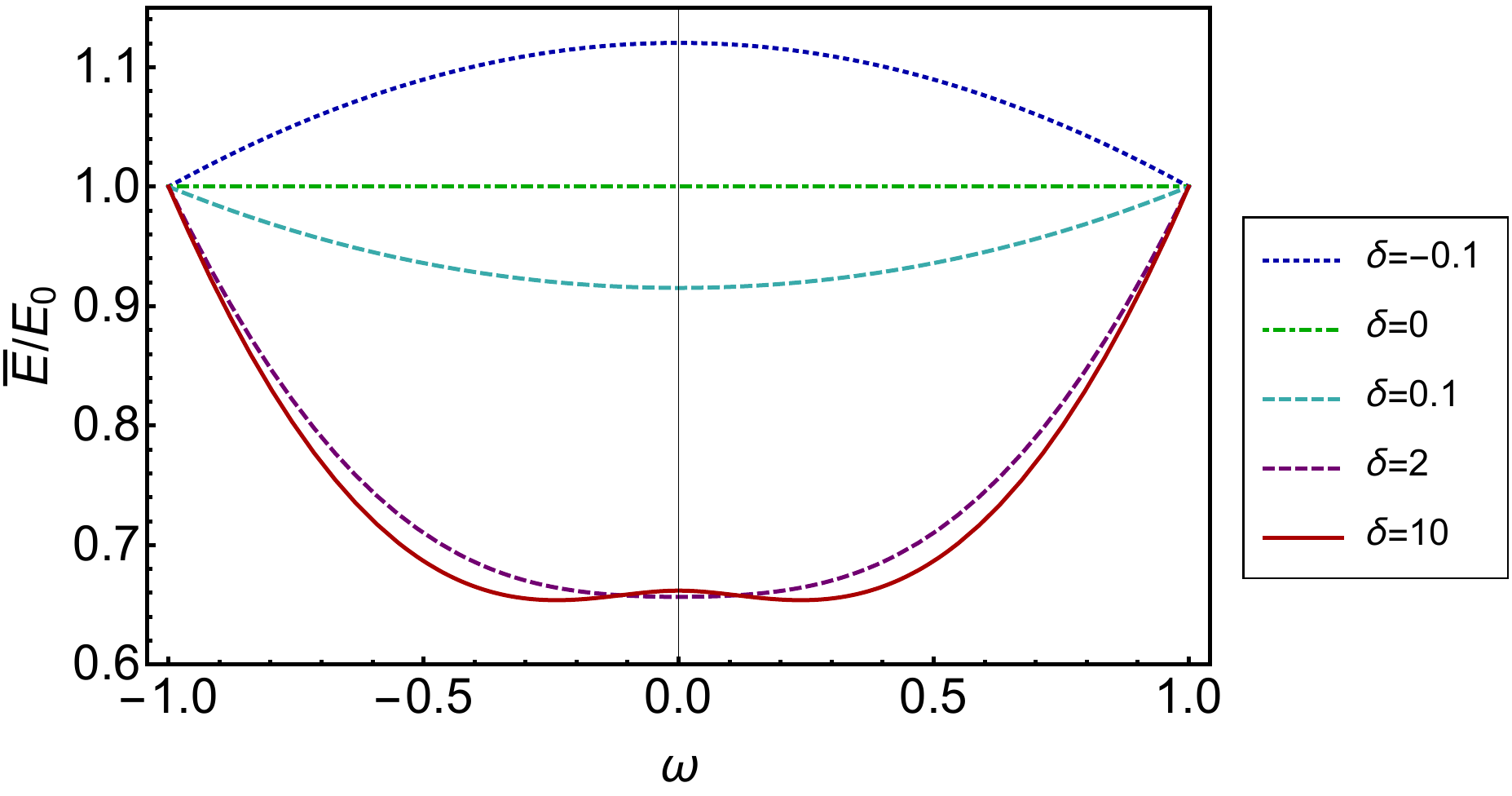}
\caption{Mean energy per unit length, $\bar{E}$, of geodesics (parametrized by the angular momentum $\omega$), for different values of $\delta$ as indicated in the figure. All oblate spheroids ($\delta<0$) are qualitatively the same - polar geodesics ($\omega=0$) have the maximal energy, while equatorial ones $\omega= \pm1$ are minimal. In the case of prolate spheroids $\delta>0$ the situation is very different: for $0<\delta<\delta^*$, $\omega_m=0$ is the minimal geodesic, while for $\delta^*<\delta$, two symmetric solutions arise $\pm \omega_m(\delta)$, which tend to the asymptotic value $\omega_{m}(\infty)= 0.255$ for extremely eccentric ellipsoids. }\label{fig:E_vs_omega}
\end{figure}

\begin{figure}[h]
	\centering
	\includegraphics[width=0.7\textwidth]{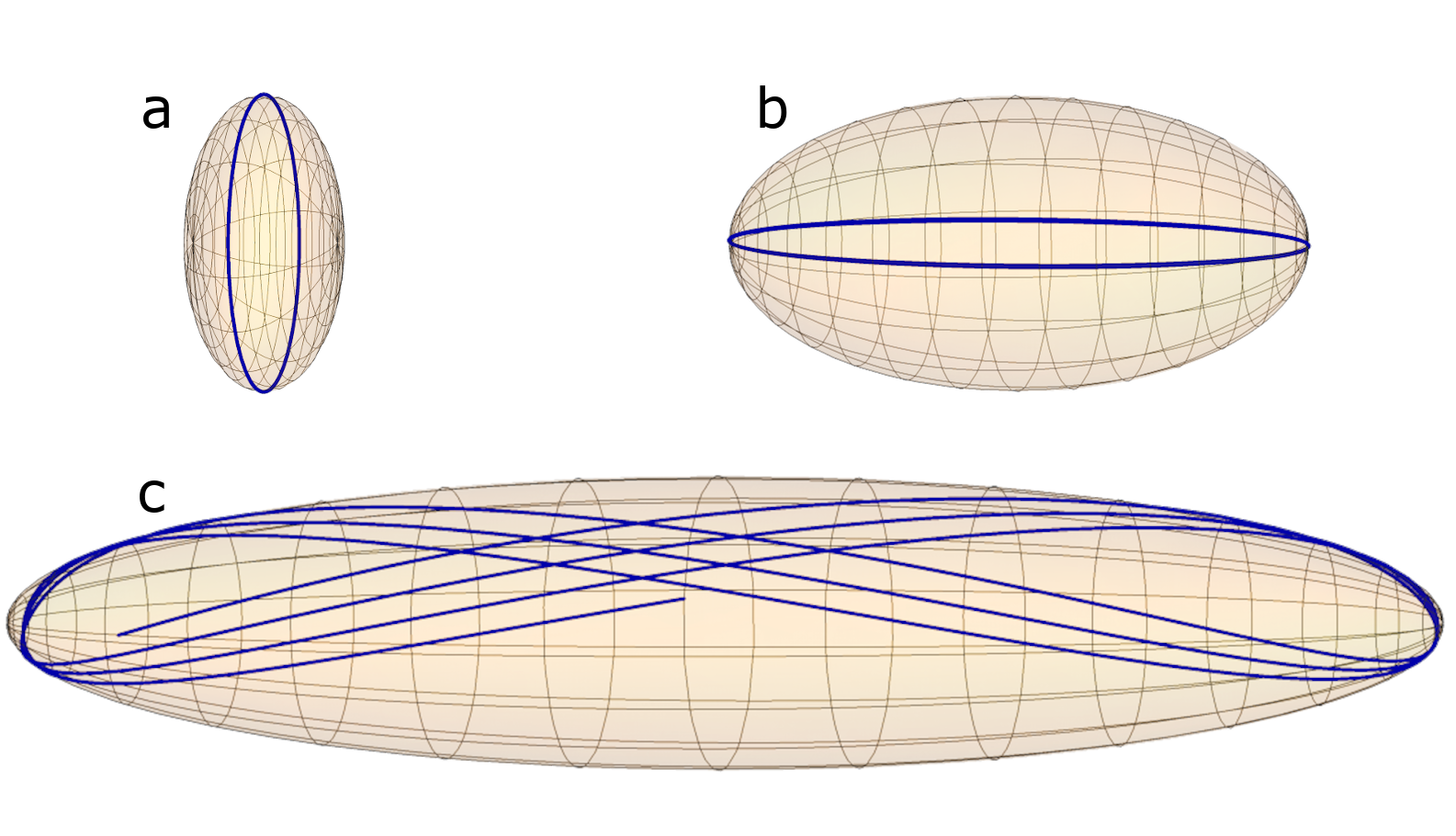}
	\caption{Visualization of  ellipsoids and their minimizing geodesics, for three different flattening ($\delta$) values. (a) $\delta =-0.5$, an equatorial geodesic is the global minimizer, (b) $\delta = 1 < \delta^*$, a polar geodesic is  the global minimizer, and  (c) $\delta =4 > \delta^*$ in which the global minimizer does not pass through the poles (only a finite portion is shown). \label{fig: minimzing geodesics} }
\end{figure}

Evaluating $\bar{E}$ on geodesics, we find that by virtue Eq. \eqref{eq:thet_geo}, 
instead of integrating over $s$ we may integrate over $\theta$ (or rather over $x$) where we find \begin{align}\frac{\dt s}{\dt x} =\sqrt{\frac{(1+\delta)^2-\delta(2+\delta)x^2}{1-\omega^2 - x^2}}.
\end{align} Thus, focusing on geodesics we look to minimize the following energy functional:
\begin{align}\label{eq:ener_glob_geo}
\frac{\bar{E}(\omega,\delta)}{E_0}&= \frac{\displaystyle\int E \dt s}{\displaystyle\int \dt s}= \frac{\displaystyle\int_{-\sqrt{1-\omega^2}}^{\sqrt{1-\omega^2}} \frac{\left(1+\delta \right)^2 \left[1+ \delta (2+ \delta ) \omega ^2\right]^2}{\left[1+\delta  (\delta +2) \left(1-x^2\right)\right]^3} \sqrt{\frac{1+(1-x^2)\delta(2+\delta)}{1-x^2-\omega^2}} \dt x }{\displaystyle\int_{-\sqrt{1-\omega^2}}^{\sqrt{1-\omega^2}}\sqrt{\frac{1+(1-x^2)\delta(2+\delta)}{1-x^2-\omega^2}} \dt x} \\
& = \frac{2\left[2+\delta(2+\delta)(1+\omega^2)\right]\mathcal{E}\left[\frac{\delta(2+\delta)(1-\omega^2)}{(1+\delta)^2}\right] - 2 \left[1+\delta(2+\delta)\omega^2\right]\mathcal{K}\left[\frac{\delta(2+\delta)(1-\omega^2)}{(1+\delta)^2}\right]}{3(1+\delta)^2\mathcal{E}\left[\frac{\delta(2+\delta)(1-\omega^2)}{(1+\delta)^2}\right]} \\ 
&= \frac{2\left[2+\delta(2+\delta)(1+\omega^2)\right]}{6(1+\delta)^2} - \frac{2\left[1+\delta(2+\delta)\omega^2\right]\mathcal{K}\left[\frac{\delta(2+\delta)(1-\omega^2)}{(1+\delta)^2}\right]}{3(1+\delta)^2\mathcal{E}\left[\frac{\delta(2+\delta)(1-\omega^2)}{(1+\delta)^2}\right]}
\end{align}
where $\mathcal{E}[m]= \int_0^{\pi/2} \sqrt{1-m \sin^2\theta} \dt \theta$ and $\mathcal{K}[m]=\int_0^{\pi/2} \left({\sqrt{1-m\sin^2\theta}}\right)^{-1} \dt \theta$ are the complete elliptic integrals of the second and first kind \cite{olver2010nist}, respectively. Plotting $\bar{E}$ for different values of  $\delta$ as a function of $\omega$ (see Fig.\ref{fig:E_vs_omega}), it is easily seen that for $\delta < 0$ only $\omega=\pm 1$ are the minima, this is true for every $-1<\delta<0$ and ,as seen in appendix A, these solution are stable. We therefore conclude that equatorial geodesics are the global minimizers for oblate spheroids (see Fig. \ref{fig: minimzing geodesics}(a)). Not surprisingly, in the case of a sphere ($\delta=0$) we find that any geodesic is a global minimizer. These results (for $\delta \leq 0$) are indeed very intuitive. In contrast, a more complex behaviour arises in the case of prolate spheroids ($\delta>0$), where we observe a sharp transition at a critical flattening $\delta^*$ (Fig. \ref{fig:E_vs_omega},  see Fig. \ref{fig: minimzing geodesics}(b) and (c) for visualization of possible curves).

\begin{figure}[!h]
\centering
\includegraphics[width=0.5\textwidth]{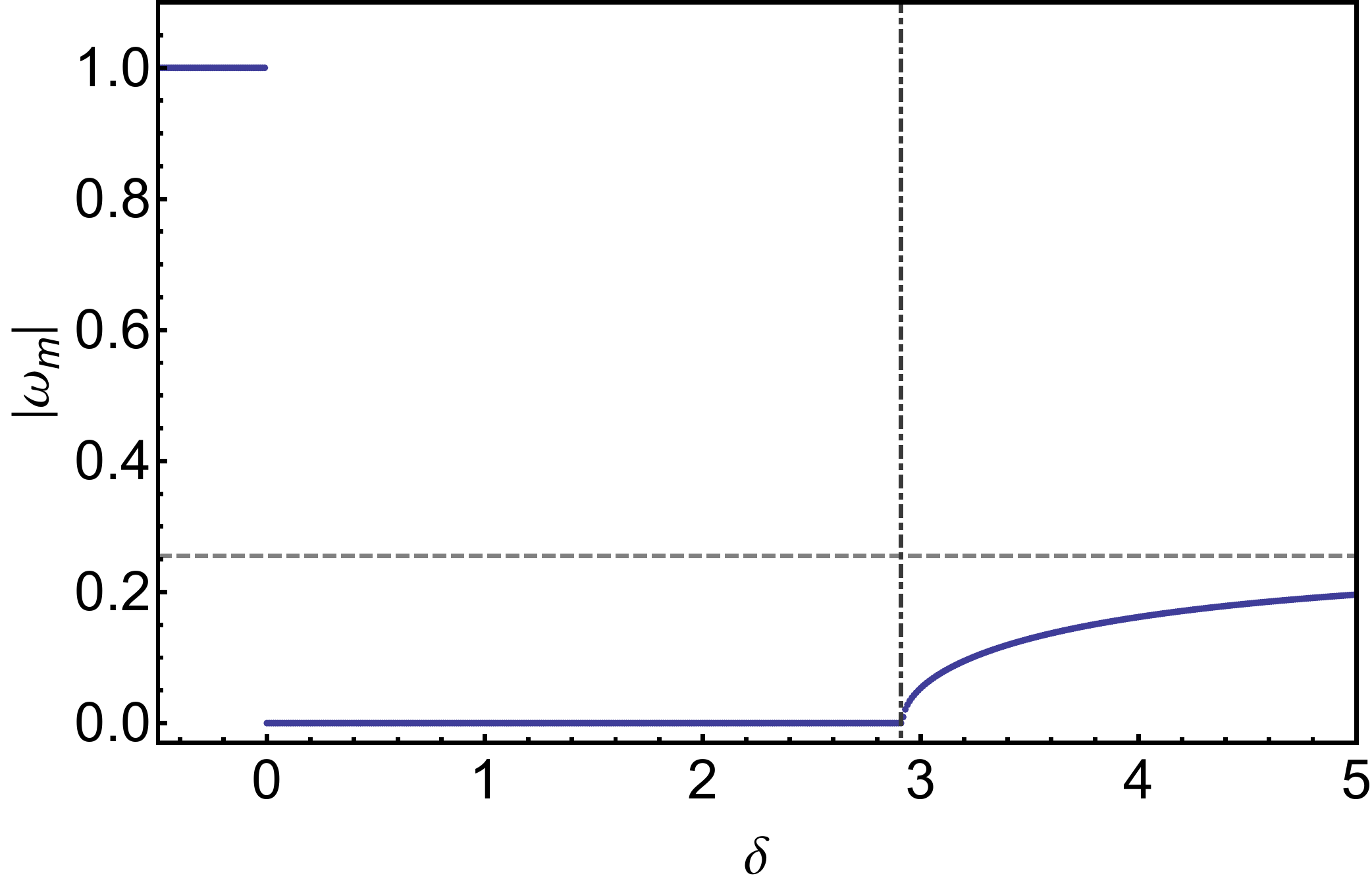}
\caption{$|\omega_m(\delta)|$ as a function of $\delta$. Note that $|\omega_m(\delta<0)|=1$. The gray dashed horizontal line is the asymptote $|\omega_{m}(\infty)| =0.255$, the vertical dot - dashed line marks the critical value $\delta^*=2.91756$.  }\label{fig:omega_min}
\end{figure}

We mark by $\omega_m(\delta)$ the minimizer of $ \bar{E}(\omega,\delta)$, and it's energy by $\bar{E}(\omega_{m}(\delta)) = \bar{E}(\omega_m (\delta),\delta)$ i.e. the globally minimizing geodesic, as a function of the flattening, $\delta$, and plot it (Fig. \ref{fig:omega_min}).  We find that for $0<\delta\leq\delta^*$ polar geodesics ($\omega_m =0$) are the minimizers of $\bar{E}$, while for  $\delta> \delta^*$, $\omega_m (\delta) \neq 0$. Solving numerically the requirement that the extremum at $\omega=0$ becomes a saddle  $\left.\frac{\pd^2 \bar{E}}{\pd \omega ^2}\right|_{\omega=0} =0$, where
\begin{align}
\left.\frac{\pd^2 \bar{E}}{\pd \omega ^2}\right|_{\omega=0} = \frac{E_0}{3} \left[5- \frac{4}{(1+\delta)^2}+ \frac{\mathcal{K}\left[\frac{\delta(2+\delta)}{(1+\delta)^2}\right]}{\mathcal{E}\left[\frac{\delta(2+\delta)}{(1+\delta)^2}\right]}\left(\frac{1}{(1+\delta)^2}\frac{\mathcal{K}\left[\frac{\delta(2+\delta)}{(1+\delta)^2}\right]}{\mathcal{E}\left[\frac{\delta(2+\delta)}{(1+\delta)^2}\right]}-2\right)\right],
\end{align}
we find that $\delta^* \simeq 2.917$. It can now be shown that near the transition, as $\delta \rightarrow \delta^{* +}$, $\omega_m (\delta) \sim \sqrt{\delta-\delta^*}$.

 In the limit of an infinitely eccentric ellipsoid  we get
\begin{align}
\frac{\bar{E}}{E_0} \xrightarrow[\delta \rightarrow \infty]{}\frac{2}{3} +\frac{\omega^2}{3}\left(2 - \frac{\mathcal{K}\left[1-\omega^2\right]}{\mathcal{E}\left[1-\omega^2\right]} \right),
\end{align}
whose minimizer is  $\omega \simeq \pm 0.255...$ This value corresponds to geodesics that intersect the equator at an angle of $\psi \simeq 75.22^\circ$ relative to the equator.

Why does this transition happen? Simply put, it is because for large $\delta$'s, the curvature at the poles becomes very large and dominant. Thus, the minimizing geodesic is such that it does not pass through the poles on one hand, yet it is still mainly directed towards the poles, as to gain as much as possible from the large difference in principal curvatures values.  This result raises the question whether there is a non-geodesic solution that has a lower mean energy even for small $\delta$, one that does not pass through the poles. Indeed, as seen in appendix B,  there must be such a solution, yet this solution is still close to a geodesic, and in any case the solution to the real physical problem must also take into account temperature, activity and self-avoidance. Our analysis  sheds light on some aspects of the mechanisms governing the actual shape and suggests that a proper treatment should take into account the surface's shape effect on the final configuration, as it serves as  an "external field" biasing the rod to align along the symmetries of the problem.

\begin{figure}[!h]
\centering
\includegraphics[width=0.5\textwidth]{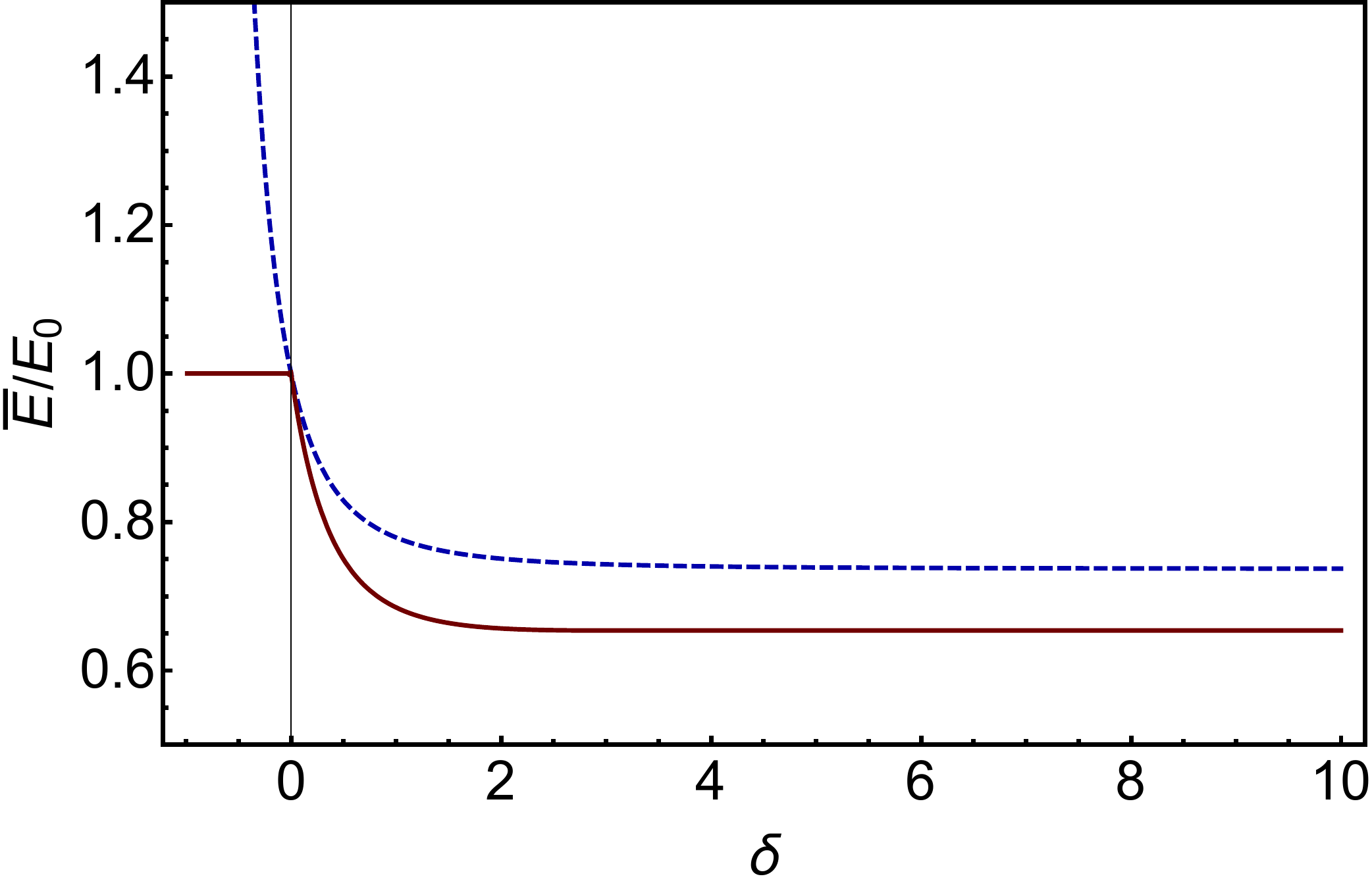}
\caption{Comparison of the energy $\bar{E}_{iso}(\delta)$ of a slowly varying isotropically oriented curve (blue dashed line) vs. the minimizing geodesic's energy $\bar{E}(\omega_{m}(\delta))$ as a fucntion of $\delta$ (solid red line).}\label{eq:ener_comp}
\end{figure}

Finally, for $\delta \neq 0$, these solutions given above, which all have a preferred direction (i.e. they are not isotropic), are clearly  preferable to any other curve that covers the ellipsoid isotropically. Any isotropic solution, that covers the ellipsoid, must have geodesic curvature, but for infinitely long filaments, one may create such a curve by slowly changing $\omega$ along a curve. In this "quasi-static" limit we change $\omega$ without introducing any geodesic curvature. Such a curve gives a lower bound on the energy of isotropic curves and is given by averaging $\bar{E}$ over all $\omega$ 
\begin{align}
\bar{E}_{iso}(\delta)= \frac{\displaystyle\int \bar{E}(\omega,\delta) \dt \omega}{\displaystyle\int \bar{E}(\omega) \dt \omega}.
\end{align} 
From Fig. \ref{eq:ener_comp} it is clearly seen that a directional (non isotropic) configuration is preferable compared to an isotropic configuration, since the dashed blue line corresponding to $\bar{E}_{iso}(\delta)$ is higher than the red solid line describing the energy of the globally minimizing geodesics $\bar{E}\left(\omega_m(\delta)\right)$.

\begin{figure}[!h]
		\centering
		\includegraphics[width=0.66\linewidth]{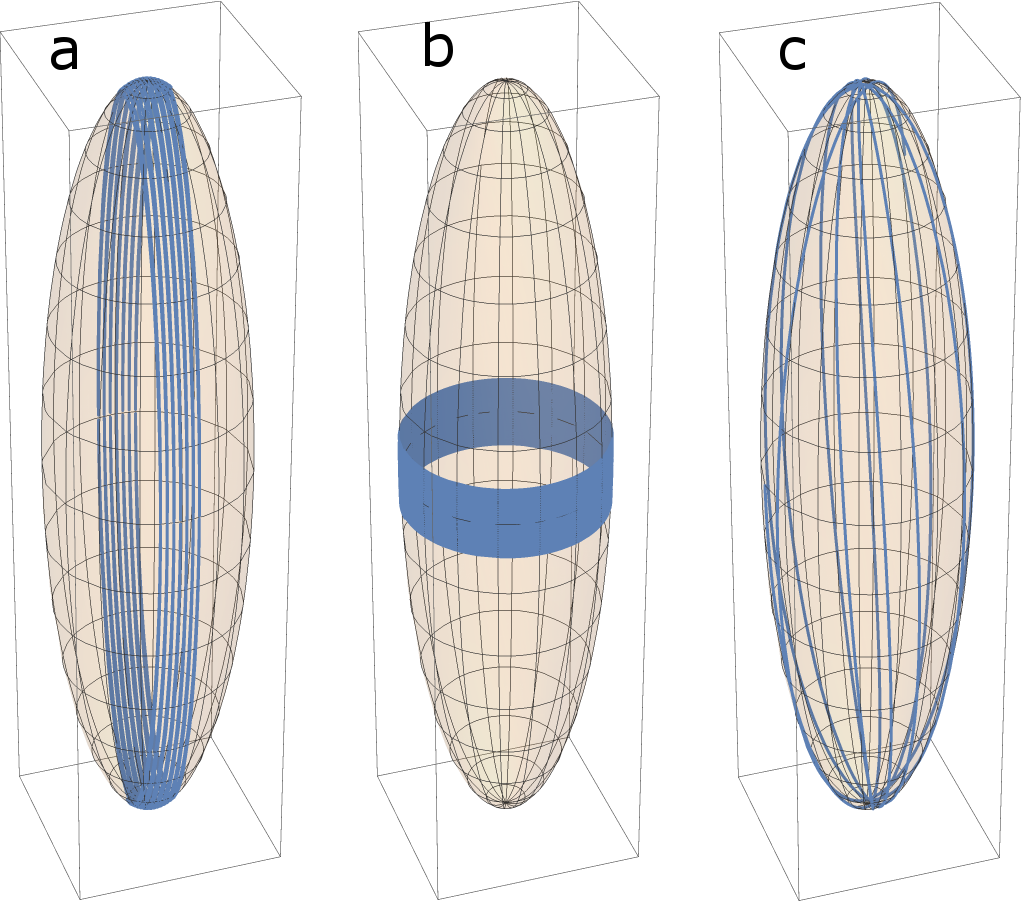}
		\caption{Three examples emphasizing effect of self avoidance on prolate spheroid packing. a) a self avoiding polar geodesic-like. b)a self avoiding equatorial geodesic - like, might be a metastable state. c) a non-self avoiding, geodesic solution, self crossing many times. }\label{fig: self avoid}
 \end{figure}

Before we conclude, a few words on the possible effect of self avoidance. Naturally, a full mechanical treatment of self avoiding filaments is hard. However we may still have a few insights using our geometrical approach. We have shown that as the ellipsoid elongates, and due to the growing curvature at the poles, a transition occurs, so that the minimizing configuration is no longer a polar geodesic. In the presence of self avoidance, this transition is likely to occur later, at higher values of $\delta$, depending on the strength of the self-avoiding interaction. The reason is that a polar geodesic - like path (see Fig. \ref{fig: self avoid} a) allows (especially for thin filaments) a non-intersecting packing, in contrast to the solution for $\delta>\delta^*$. The energy associated with a self avoiding (SA) packing is (see appendix \ref{appndx: self avoidance})
\begin{align}\label{eq:E_SA}
\frac{\bar{E}_{SA}(c,\delta)}{E_0} &=  \frac{2 \arcsin (c)}{c \sqrt{1- c^2}+\arcsin(c)}  \times \left(\frac{2\left(2+\delta(2+\delta)\right)}{3\left(1+\delta \right)^2} - \frac{\mathcal{K}\left[\frac{\delta(2+\delta)}{(1+\delta)^2}\right]}{3(1+\delta)^2\mathcal{E}\left[\frac{\delta(2+\delta)}{(1+\delta)^2}\right]} \right) \\ \nonumber
&= \frac{2 \arcsin (c)}{c \sqrt{1- c^2}+\arcsin(c)}  \times \bar{E}(\omega=0,\delta)
\end{align}
 where $1 \geq c = \frac{t L}{A}=\frac{tL}{2\pi} \frac{\sqrt{\delta\left(2+\delta\right)}}{\sqrt{\delta\left(2+\delta\right)}+ \left(1+\delta\right)^2 \text{sec}^{-1}\left(1+\delta\right)}$ is the average surface density of the filament, and $t$ is the thickness of te filament. 
 Nevertheless, when the flattening , $\delta$, is large enough, it is clear that even a self-avoiding system  will prefer to minimize the bending energy by avoiding the poles. Since any self - intersecting configuration, must do so many times, proportionate to the filament density $c$, and by denoting the self - interaction energy per unit length in $u$ , we can express  the energy  of a self - crossing filament as $\bar{E} + u c$ ($u>0$ for self - avoidance). Thus, the condition of a self - interacting filament to change configuration from polar to a non-polar one is  
 \begin{align}\label{eq: self_avoidance}
\bar{E}_{SA}(c,{\delta}) \gtrsim  \bar{E} \left({\omega_m}({\delta})\right) + u c,
 \end{align} as this expression means that the energy of a non - intersecting, self - avoiding curve is larger than that of a self interacting curve, including the self- interaction term. At small $c$'s, $\bar{E}_{SA}(c,\delta) \simeq \bar{E}(\omega =0,\delta)\left(1+\frac{c^2}{3}\right)$. Expanding $\delta^* = \delta ^*_0 + \delta^*_1 c$, where $\delta^*_0 = 2.917$ is the critical $\delta$ found earlier, and  requiring equality in Eq. \eqref{eq: self_avoidance}, using Eq. \eqref{eq:E_SA} gives us the dependence of $\delta^*$ on $c$
 \begin{align}
\bar{E}(0,\delta^*_0 +c \delta^*_1)\left(1+\frac{c^2}{3}\right)& = \bar{E} \left({\omega_m}(\delta^*_0+c \delta^*_1)\right) + u c, 
 \end{align}
 Expanding in orders of $c$ and keeping only leading orders, the above equation reduces to
 \begin{align}
 c \delta^*_1 \left.\frac{\pd \bar{E}(0,\delta)}{\pd \delta}\right|_{\delta\rightarrow {\delta^*_0}^{+}} &=u c +  c\delta^*_1 \left.\frac{\pd \bar{E}\left({\omega_m}(\delta)\right)}{\pd \delta}\right|_{\delta\rightarrow {\delta^*_0}^{+}} .
\end{align}
This can then be solved  for $\delta^*_1$
\begin{align}
\delta^*_1 &= \left.\frac{u}{\frac{\pd \bar{E}(0,\delta)}{\pd \delta}-\frac{\pd \bar{E}\left({\omega_m}(\delta)\right)}{\pd \delta}} \right|_{\delta\rightarrow {\delta^*_0}^{+}}.
 \end{align}
Since $\frac{\pd \bar{E}(0,\delta)}{\pd \delta}|_{\delta\rightarrow {\delta^*_0}^{+}} \geq 0$ and $\frac{\pd \bar{E}\left({\omega_m}(\delta)\right)}{\pd \delta}|_{\delta\rightarrow {\delta^*_0}^{+}} <0 $, it is clear that $\delta^*_1>0$ therefore self avoidance indeed postpones the aforementioned transition.

 Another possible effect of self avoidance is the appearance of meta-stable states, where the filament packs without crossing, such as paths with $\omega \sim 1$ that wrap many times around the ellipsoid (see Fig. \ref{fig: self avoid} b), but any deviation from this configuration means that the filament must cross itself many times (Fig. \ref{fig: self avoid} c). At high densities and strong self avoiding interactions	, such a configuration is likely to be meta-stable.

 To conclude, we have used a geometrical approach to show that the shape of an ellipsoid  acts as an "external field" that orders the orientation of a stiff filaments on it's surface. We have shown that energy, not only entropy, plays a significant role in ordering the packing of filaments inside non-spherical volumes. By the use of geodesics to allow for a simple  analytical treatment, we have directly shown that in general, filaments will tend to align with the long dimension of the ellipsoid. However, this is not completely accurate, since (at least for geodesics) the transition is likely to occur in very long  ellipsoids, where the globally bending minimizing  curve no longer passes through the poles. The transition is somewhat sensitive to the geometry,e.g. within a model of cylinder with spherical caps, no such transition will occur as a function of elongation as the curvature does not depend on the the flattening in such a case.  Nevertheless, these results reflect similar numerical result as in \cite{mirabet2018self,vetter2013finite}, specifically, the transition described in this paper echoes numerical results found in the context of microtubules \cite{marenduzzo2008computer,marenduzzo2009statistics,ali2006polymer,jun2006entropy}, and the folded toroid - to - twisted toroid transition described by \cite{petrov2007conformation,petrov2008packaging,petrov2009characterization} in the context of viral capsids. Additionally, most wild type viral capsids have a relatively small flattening, which is below the transition found here, namely $\delta < \delta^*$. The results shown here suggest that a too eccentric shape will  make packing of genetic material not as efficient with respect to injection into a host cell.  When applied to the ordering of MT along cell walls, these results suggest that ordering does not necessitate any biological means of actively or passively sensing stress within the cell wall. Hence, sensitivity to stress/strain  by proxi, namely the intgeral over strain, leads to anisotropy. 
 Finally, the treatment shown here is simplified, nor does not fully take into considerations any entropic effect and geometric effects such as self avoidance. In any real system, temperature and/or activity  play a significant role, as well as self avoidance \cite{edwards1965statistical,katzav2006statistical,boue2007folding}, and should be taken into account. Additionally, dynamics \cite{ali2006polymer}, friction  effects \cite{vetter2014morphogenesis} (both dynamic and static), plasticity of either the membrane or the filaments \cite{vetter2014morphogenesis}, and electrostatics \cite{marenduzzo2008computer}, were not taken into account and are likely to play roles in different systems. In this context, the results shown here are an important stepping stone. A full statistical mechanical treatment that takes into account both entropy and self avoidance is postponed for a later paper.
\newline
The authors would like to thank Enrico Cohen for his useful remarks and discussion.

\clearpage
\appendix

\section{Stability Analysis}\label{appndx: stability}

For  stability analysis we use a somewhat different and much more technical approach than the one presented in the main text. Most notably, we use an index - less notation, since an index notation will be much more cumbersome and definitely more complex to read. The analysis presented here is based on standard differential geometry as can be found, in example in \cite{do2016differential}. It begins by rewriting our energy expression (Eq. \eqref{eq:rod_ener}) for a general curve, parametrized generally (Eq. \eqref{eq:ener_glob_diffgeo}), that is not necessarily using the natural parametrization of arc-length $|R'(s)|=1$.  Soon after, we rigorously define a variation of the curve on a differential manifold, which allows us to define a variational field, which is the infinitesimal of a variation about the curve. This allows us to calculate the first and, more importantly, the second variations of our energy functional.

We begin by marking a general curve $c(s)$. The tangent to that curve is $\bar{\nabla}_s c=\frac{D c}{D s} = \frac{\partial c}{\partial s}=\dot{c}$. The energy we are seeking to minimize is then 
\begin{align}\label{eq:ener_glob_diffgeo}
	\bar{E}&=  \frac{E}{L}= \frac{\displaystyle\int  \left (\frac{\langle \bar{\nabla}_s \dot{c},\bar{\nabla}_s \dot{c}\rangle }{\langle\dot{c},\dot{c} \rangle^2}  - \frac{\langle \nablab_s\dot c, \dot c \rangle^2}{\langle \dot c, \dot c \rangle^3} \right)\sqrt{\langle \dot{c},\dot{c}\rangle} \dt s  }{\displaystyle\int\sqrt{\langle \dot{c},\dot{c}\rangle} \dt s } = \frac{\displaystyle\int \frac{\langle \bar{\nabla}_s \dot{c},\bar{\nabla}_s \dot{c}\rangle }{\langle\dot{c},\dot{c} \rangle^{3/2}} \dt s  }{\displaystyle\int\sqrt{\langle \dot{c},\dot{c}\rangle} \dt s },
	\end{align}
	 where we omitted the second term in the first integral since it will prove to contribute exactly nothing in the following calculations.  $\bar{\nabla}$ is the connection in the ambient space ($\mathbb{R}^3$), $\langle X ,Y \rangle$ is the inner product (metric) of two vectors $X,Y$. $\dot{c} \equiv \frac{\pd c}{\pd s} = \bar{\nabla}_s c $ since $c$ is a scalar defined on the ellipsoid. Note that this form of writing is almost identical to Eq.\eqref{eq:rod_ener}, with the difference that we may parametrize the curve in any way we want, therefore we include in the expression of  $\langle \bar{\nabla}_s \dot{c},\bar{\nabla}_s \dot{c} \rangle$ a tangent term $\langle \dot{c}, \bar{\nabla}_s\dot{c}\rangle$ and a normalization $\langle \dot{c},\dot{c} \rangle$. Since eventually we will use arc-length parametrized curves (in fact, geodesics), this term will vanish.

	A variation, $h(s,t)$, of a curve $c(s)$, is a two parameter function ($s \in [0,L], t \in (-\epsilon, \epsilon)$), such that $h(s,0)= c(s)$, and $h(0,t)=c(0), h(L,t) =c(L)$. $h(s,t)\equiv c_t(s)$ for a given $t$ is some curve meandering about $c(s)$.  The field $v(s)=\frac{\pd h}{\pd t}(s,0)$ is called the variational field. Furthermore, close to $c(s)$, one can write $h(s,t)$ locally, without loss of generality , as $h(s,t) = \exp_{c(s)}(t v(s))$. Where $\exp_p(v)\equiv \gamma(p,1,v)$ is the exponential map about a point $p$ (a step of "length" $1$ along a geodesic $\gamma(p,\lambda,v)$, so that  $\gamma(p,0,v)=p, \dot{\gamma}(p,0,v) = \frac{\pd \gamma(p,\lambda,v)}{\pd \lambda}_{\lambda=0}=v$). Finally, the $n^{th}$ variation  of a functional $F[c(s)]$ is then found by $\delta^n F = \frac{1}{n!}\frac{d^n F[h(s,t)]}{d t^n}|_{t=0}$. 
	
	Thus
	\begin{align}
	\bar{E}(t)= \frac{E(t)}{L(t)}= \frac{\displaystyle\int \frac{\langle \bar{\nabla}_s \dot{h},\bar{\nabla}_s \dot{h}\rangle }{\langle\dot{h},\dot{h} \rangle^{3/2}} \dt s  }{\displaystyle\int\sqrt{\langle \dot{h},\dot{h}\rangle} \dt s },
	\end{align}
	and we look for extrema by requiring a vanishing first variation $\delta \bar{E}$ 
	\begin{align}\label{eq:var_E}
	\delta \bar{E}= \frac{\delta E}{L}- \frac{E}{L}\frac{\delta L}{L}= \frac{\delta E}{L}- \bar{E}\frac{\delta L}{L} =0,
	\end{align}
	so that $\delta L= 0$ (i.e. curves are geodesics), and $\delta E = 0$ (they are extrema of the energy). In this case it is easy to show that the second variation is then
	\begin{align}\label{eq:2nd_var_E}
	\delta^2 \bar{E}= \frac{\delta^2 E}{L}- \bar{E}\frac{\delta^2 L}{L}.
	\end{align}	
	Since $L$ is the length of a curve, its minimizers are geodesics:
		\begin{align}\label{eq:geo_var_1}
		2\delta L &= 2\displaystyle\int \frac{d}{dt} \sqrt{ \langle \dot{h}, \dot{h} \rangle}  \dt s = \int \frac{\langle \bar{\nabla}_t  \bar{\nabla}_s h , \bar{\nabla}_s h \rangle}{\sqrt{\langle \dot h, \dot h \rangle}} \dt s  \\ \nonumber
		&\stackrel{t=0}{=} \int \frac{\langle  \bar{\nabla}_s v ,\dot{c} \rangle}{\sqrt{\langle \dot{c}, \dot{c}\rangle}} \dt s= \int \frac{d}{ds} \left(\frac{\langle  v ,\dot{c} \rangle}{\sqrt{\langle \dot{c}, \dot{c}\rangle}} \right) \dt s - \int  \frac{\langle  v , \nablab_s\dot{c} \rangle}{\sqrt{\langle \dot{c}, \dot{c}\rangle}} -  \frac{\langle  v ,\dot{c} \rangle\langle  \dot{c} , \nablab_s\dot{c} \rangle}{\sqrt{\langle \dot{c}, \dot{c}\rangle}^3} \dt s
		\end{align}
	The first term is identically zero as it is a full derivative term and $v=0$ at the boundary (even when they are taken to infinity). As for the second term, we note that it vanishes iff $c(s) = \gamma(s)$ is  a geodesic. In that case $\langle \dot \gamma, \nablab_s \dot \gamma \rangle=0$ is immediately satisfied ( by the very definition of a geodesic), and  $\nablab_s \dot \gamma= \Bb{\dot \gamma}{\dot \gamma} \hat{n}$, where $\Bb{X}{Y} = \langle \bar{\nabla}_Y X , \hat{n} \rangle$ is the second fundamental form and $\hat{n}$ is the surface normal. Since $v \perp \hat{n}$ (it is a derivative of a scalar hence it  is tangent to the surface), we immediately get in such a case that $\langle v, \nablab_s \dot \gamma \rangle =0$	
	
	Thus geodesics are extremal curves if  
	\begin{align}
		0&=\delta E=\int  \delta \langle \bar{\nabla}_s \dot{h},\bar{\nabla}_s \dot{h}\rangle  -\frac{3}{2}  \langle \bar{\nabla}_s \dot{\gamma},\bar{\nabla}_s \dot{\gamma}\rangle \delta \langle \dot{h},\dot{h}\rangle \dt s |_{t=0}  = \int  \delta \langle \bar{\nabla}_s \dot{h},\bar{\nabla}_s \dot{h}\rangle ,
	\end{align}
	where we used the fact that we are on geodesics hence $ \delta \langle \dot{h},\dot{h}\rangle $ is zero as just has been verified. It is readily seen that
	\begin{align}
	\int \delta \langle \bar{\nabla}_s \dot{h},\bar{\nabla}_s \dot{h}\rangle \dt s &= \int \dt s \frac{d}{d t} \langle \bar{\nabla}_s \dot{h},\bar{\nabla}_s \dot{h}\rangle |_{t=0} =2 \int \dt s  \, \langle \nablab_t \nablab_s \dot{h}, \nablab_s \dot h \rangle = 2 \int \dt s  \, \langle \nablab_s^2 v, \nablab_s \dot \gamma \rangle \\ \nonumber
	&= 2 \int \dt s \, \Bb{\dot{\gamma}}{\dot{\gamma}} \langle \nablab_s^2 v,\hat{n} \rangle,
	\end{align}
	where we used the fact that we are embedded in flat Euclidean space, hence that covariant derivatives, $\nablab$, commute, as the Riemann tensor vanishes.  Using the fact that  one can decompose $\bar{\nabla}= \nabla + \nabla^\perp$, where $\nabla$  ($\equiv\nabla^\top$)is the connection on the tangent bundle of our ellipsoid, and $\nabla^\perp$ is the normal connection, 
	\begin{align}
	\langle \nablab_s^2 v, \hat{n} \rangle &= \langle \nablab_s \left( \nabla_s^\bot v + \nabla^\top_s v\right), \hat{n} \rangle = \langle 
	{\nabla_s^\top}^2 v+ \nabla_s^\bot \nabla_s^\top v +\nablab_s \nabla_s^\bot v ,\hat{n} \rangle. \\ 
	\intertext{The first term on the right hand side vanishes, since by defition the connection on the tangent bundel vanishes on the normal bundle - $\langle \nabla^\top_Y X, \hat{n} \rangle=0$ , and $ \frac{d}{d s}\Bb{X}{Y}  = \bar{\nabla}_s \langle \bar{\nabla}_Y X,  \hat{n}\rangle =\langle \bar{\nabla}_s \bar{\nabla}_Y X,  \hat{n}\rangle + \langle\bar{\nabla}_Y X,  \bar{\nabla}_s \hat{n}\rangle$. Additionally, $\bar{\nabla}_X \hat{n} = \nabla^\top_X \hat{n}$ for any $X$ by virtue of $\hat{n}$ being a unit vector. We thus get}
	\langle \nablab_s^2 v, \hat{n} \rangle &=\Bb{\nabla_s^\top v}{\dot{\gamma}} + \frac{d}{d s}\Bb{v}{\dot{\gamma}},  \\ 
	\intertext{and from the definition of the covariant derivative of $\mathbf{B}$, together with the definition of a geodesic curv $\nabla_s \dot{\gamma} = 0$,}
	\langle \nablab_s^2 v, \hat{n} \rangle&=\Bb{\nabla_s v}{\dot{\gamma}} + \left(\nabla_s^\bot \mathbf B\right) \left(v,\dot{\gamma}\right) + \Bb{\nabla_s v}{\dot{\gamma}} + \Bb{v}{\nabla_s\dot\gamma} \\ 
	&= \left(\nablab_s \mathbf B\right) \left(v,\dot{\gamma}\right) +2 \Bb{\nabla_s v}{\dot{\gamma}}.
	\end{align}

	Thus, $\delta E$ is zero iff either $\gamma$ is an asymptotic curve (in which case $\Bb{\dot{\gamma}}{\dot{\gamma}}$ is zero), or if $\gamma$ is locally a principal curve, that is tangent to the principal directions of the $\mathbf{B}$. The latter can be seen by noting that around a geodesic, we may restrict ourselves to variational fields $v$ that are perpendicular to the curve without any loss of generality ,as any tangent component of $v$ translates into re-parametrization of the geodesic). As a result, if $\gamma$ is also a principal curve, then $v$ always points in the other principal direction, and so is $\nabla_s v$ as can directly be seen by taking the derivative along the geodesic of $\langle v, \dot{\gamma}\rangle =0$. Therefore $\Bb{v}{\dot{\gamma}}$ and $\Bb{\nablab_s v}{\dot\gamma}$ are identically zero.
	
	To summarize - geodesics that are either asymptotic curves or principal curves, are extrema of $\bar{E}$. In order to analyse their stability, we need to calculate the second variation. It is easily shown that around a geodesic curve
	\begin{align}
	L \delta^2 \bar{E} = \int \dt s \left[ \delta^2 \langle \nablab_s \dot h ,\nablab_s \dot h \rangle  - \frac{1}{2} \left( 3\langle \nablab_s \dot\gamma, \nablab_s \dot\gamma \rangle +\bar{E} \right) \delta^2 \langle \dot h, \dot h \rangle  \right].
	\end{align}
	Note that around our extrema
	\begin{align}
	\delta^2 \langle \nablab_s \dot h ,\nablab_s \dot h \rangle  &= \frac{1}{2} \frac{d^2}{d t^2} \langle \nablab_s \dot h ,\nablab_s \dot h \rangle 
	\end{align}
	Hence 
	
	\begin{align}
	2 L \delta^2 \bar{E} = \int \dt s \left[ \frac{d^2}{d t^2} \langle \nablab_s \dot h ,\nablab_s \dot h \rangle - \frac{1}{2} \left( 3\Bb{\dot{\gamma}}{\dot{\gamma}}^2 +\bar{E} \right) \frac{d^2}{d t^2}\langle \dot h, \dot h \rangle  \right].
	\end{align}

	Now \cite{do2016differential}
	\begin{align}
	\frac{1}{2}\frac{d^2}{d t^2} \langle \dot{h} ,\dot{h} \rangle= \langle \nabla_s v ,\nabla_s v \rangle - K \langle v, v \rangle
	\end{align}
	where $K$ is the Gaussian curvature.
	
	Therefore
	\begin{align}
	\frac{1}{2} \frac{d^2}{d t^2} \langle \nablab_s \dot h, \nablab_s \dot h \rangle &= \frac{d}{dt} \langle \nablab_s^2 \pd_t h,\nablab_s  \dot h \rangle = \langle \nablab_s^2 \nablab_t \pd_t h, \nablab_s \dot h \rangle + \langle \nablab_s^2 v, \nablab_s^2 v \rangle.
	\end{align}	
	Calculating it term by term (starting with the first), using the fact that $v$ is a tangent to a geodesic along $t$,hence $\nablab_t v$ is in the normal bundle-
	\begin{align}
	\langle \nablab_s^2 \nablab_t \pd_t h, \nablab_s \dot h \rangle &= \langle \nablab_s^2 \left( \Bb{v}{v} \hat{n}\right), \Bb{\dot{\gamma}}{\dot{\gamma}} \hat{n} \rangle = \Bb{\dot{\gamma}}{\dot{\gamma}} \langle \nabla_s^\bot \nablab_s \left(  \Bb{v}{v} \hat{n} \right) ,\hat{n} \rangle.
	\end{align}
	A direct expansion yields
	\begin{align}
	\nablab_s \left( \Bb{v}{v} \hat{n} \right)&	= \frac{d}{ds} \Bb{v}{v} \hat{n} + \Bb{v}{v} \left(-S_{\hat{n}} \dot{\gamma}\right) =  (\nablab_s \mathbf{B})(v,v) \hat {n} + 2 \Bb{\nabla_s^\top v}{v} \hat{n} - \Bb{v}{v}S_{\hat{n}} \dot{\gamma},
	\end{align}
	where $S_{\hat{n}} x =-\nabla_x \hat{n}= -\sum \limits_{i=1,2} \langle \nabla_x \hat{n}, e_i \rangle e_i$ is the shape operator along $x$ ($e_i$ are an orthonormal frame on the surface). It is immediate  that 
	\begin{align}
	\nabla_s^ \bot \nablab_s \left(\Bb{v}{v} \hat{n}\right) &=
	\frac{d}{d s}(\nablab_s \mathbf{B})(v,v) \hat {n} + 2  \frac{d}{d s}\Bb{\nabla_s^\top v}{v} \hat{n} - \Bb{v}{v} \nabla_s^\bot \left( S_{\hat{n}}  \dot{\gamma} \right).
	\end{align}
	From the shape operator definition, it is clear that
	\begin{align}
	\nabla_s^\bot \left( S_{\hat{n}}  \dot{\gamma} \right) &= - \nabla_s^\bot \sum_i  \langle \nabla_s \hat{n},e_i \rangle e_i = \nabla_s^\bot \sum_i  \Bb{\dot{\gamma}}{e_i} e_i = \sum_i \Bbs{\dot{\gamma}}{e_i} \hat {n}.
	\end{align}
	We thus conclude that 
	\begin{align}
	\langle \nablab_s^2 \nablab_t \pd_t h, \nablab_s \dot h \rangle |_{t=0} &=  \Bb{\dot{\gamma}}{\dot{\gamma}} \left[ \frac{d}{d s}(\nablab_s \mathbf{B})(v,v) + 2  \frac{d}{d s}\Bb{\nabla_s^\top v}{v} -\sum_i\Bb{v}{v} \Bbs{\dot{\gamma}}{e_i} \right] \\ \nonumber
	&=  \Bb{\dot{\gamma}}{\dot{\gamma}} \left[ \frac{d}{d s}(\nablab_s \mathbf{B})(v,v) + 2  \frac{d}{d s}\Bb{\nabla_s^\top v}{v} \right]- K\, \Bbs{\dot{\gamma}}{\dot{\gamma}} \langle v,v \rangle
	\end{align}
	where we used the fact that $\gamma$ is a geodesic which is also a principal curve, and that $\langle v, \dot{\gamma}\rangle =0$. Thus $\Bb{\dot{\gamma}}{e_i} = k_\gamma \sqrt{\langle \dot{\gamma}, \dot{\gamma} \rangle}$ if $e_i$ points along $\dot{\gamma}$ and $0$ otherwise. $k_\gamma$ is the principal curvature along $\gamma$. Similarly $ \Bb{v}{v}= k_v \langle v,v\rangle = k_v \|v\|^2$. Finally, $K= k_\gamma k_v$.
	
	We thus conclude that
	\begin{align}
		L \delta^2 \bar{E} = \int \dt s & \left\{  \left(\|\nablab_s^2 v\|^2 + \Bb{\dot{\gamma}}{\dot{\gamma}} \left[ \frac{d}{d s}(\nablab_s \mathbf{B})(v,v) + 2  \frac{d}{d s}\Bb{\nabla_s^\top v}{v} \right]- K\, \Bbs{\dot{\gamma}}{\dot{\gamma}} \| v \|^2 \right)\right. \\ \nonumber & \left.  - \frac{1}{2} \left( 3\Bbs{\dot{\gamma}}{\dot{\gamma}} +\bar{E} \right)  \left(\|\nabla_s v\|^2 - K \| v\|^2\right) \right\} \\ \nonumber
		 = \int \dt s & \left\{ \left\| \nablab_s^2 v \right\|^2 - \frac{1}{2}\left(3\Bbs{\dot{\gamma}}{\dot{\gamma}} +\bar{E} \right)  \left\| \nablab_s v \right\|^2 + \frac{K}{2}\left(\Bbs{\dot{\gamma}}{\dot{\gamma}} +\bar{E} \right)  \left\|  v \right\|^2  + \Bb{\dot{\gamma}}{\dot{\gamma}} \frac{d^2}{d s^2} \Bb{v}{v}  \right\} \\ \nonumber
		 \intertext{Integrating by parts the last expression twice yields}
		 	L \delta^2 \bar{E}= \int \dt s & \left\{ \left\| \nablab_s^2 v \right\|^2 - \frac{1}{2}\left(3\Bbs{\dot{\gamma}}{\dot{\gamma}} +\bar{E} \right)  \left\| \nablab_s v \right\|^2 + \frac{K}{2}\left(\Bbs{\dot{\gamma}}{\dot{\gamma}} +\bar{E} \right)  \left\|  v \right\|^2  +  \Bb{v}{v} \frac{d^2}{d s^2}  \Bb{\dot{\gamma}}{\dot{\gamma}} \right\}.		 
	\end{align}

It is  immediately seen that on a sphere (by setting $K=1$, $\Bb{X}{Y}=1$)

\begin{align}
		L \delta^2 \bar{E}_{sphere} = \int \dt s & \left( \left\| \nablab_s^2 v \right\|^2 - 2  \left\| \nablab_s v \right\|^2 +   \left\|  v \right\|^2 \right).
	\end{align}
Using Fourier analysis, it is easily shown that this expression is stable with respect to all variational fields $v$ except $v= v_0 \exp(i s)$, where it vanishes. This means that every variation around a geodesic (which is, by default, a principal direction) is stable (i.e - geodesics are stable solutions) except ones which oscillate with a periodicity of the sphere's circumference. Such a variation is merely another (nearby) great circle (all and only great circles  are geodesics on a sphere).

By expressing $\Bb{X}{Y}$ for polar geodesics ($\omega =0$), and expanding the expressions in small $\delta>0$ it can be shown that instability occurs for wave-numbers (around polar geodesics)
\begin{align}
1-\sqrt{2 \delta } <\lambda < 1+ \sqrt{2 \delta},
\end{align}
hence polar geodesics are unstable for all $\delta>0$.
For $\delta<0$ all, equatorial geodesics are always stable since 
\begin{align}
L \delta^2 \bar{E}\left({\delta<0}\right) \simeq \int \dt s &\left( \left\| \nablab_s^2 v \right\|^2 - 2  \left\| \nablab_s v \right\|^2 +   \left\|  v \right\|^2 + 2|\delta|  \left\|  v \right\|^2 \right)  >0 .
\end{align}

\section{Estimation of Stable solutions} \label{appndx: deviation estimation}

While one can produce better estimates, here we will use only a rough estimation to show that for $0<\delta \ll 1$, $\kappa_g$ must be small too. The logic behind this argument is as follows - it is clear that a curve on the equator will, at least locally, prefer to point along the polar direction as the curvature is lowest along this direction. However near the poles the curvature is larger even than that of the equatorial geodesic, it may therefore be preferential for the curve to change direction so that it bypasses the pole, even if only slightly.  In order to bound the average geodesic curvature, we consider two cases: as an upper bound we consider a geodesic on a sphere having the  same curvature as that at the pole ($l_{\max}= 1+\delta$). Clearly, any physical solution on the ellipsoid has  lower energy than that, since even a polar geodesic has lower energy. As for a lower bound we now consider some configuration (with non-zero geodesic curvature) on a sphere whose curvature equals to the lowest curvature on the ellipsoid ($l_{min}=\frac{1}{(1+\delta)^2}$). In other words
\begin{align}
\bar{E}_{lower} = \langle \kappa_g^2 \rangle +l^2_{min} < \bar{E}_{actual} =\langle \kappa_g^2 \rangle &+\langle l^2(\kappa_g) \rangle  < \bar{E}_{higher} = l^2_{max}  \\ \nonumber   
\Rightarrow \langle \kappa_g^2\rangle <  l^2_{max} -l^2_{min} =  (1+\delta)^2 &- 1/(1+\delta)^4 \stackrel{\delta \ll 1}{\simeq} = 6\delta + O(\delta^2) \ll 1.
\end{align}
This is an extremely rough estimate, but it suffices in order to support our claim that for small $\delta$ the actual minimizing curve is almost a geodesic (and specifically a polar geodesic). Note that a localized highly curved region is not a possible solution since for any finite deviation from a geodesic assuming a localized curvature  along some scale $\Delta$, one must have $\kappa_g \propto 1/\Delta$. Furthermore, from the symmetry of the problem, such regions must occur infinitely many times along the curve  (if it occurs only a finite number of times, than we must sit on a geodesic). Hence the contribution to the mean energy $\Delta \bar{E} \propto 1/\Delta^2 \Delta=1/\Delta \xrightarrow{\Delta \rightarrow 0}\infty$. Hence the globally minimizing curve must have a fairly uniform distribution of geodesic curvature along it (i.e. $\kappa_g \sim const. \sim 2 \sqrt{\delta} \ll 1$)

In the case of $\delta \gg 1$ a different argument takes over - in this limit, the polar direction has mostly zero curvature, except at a region near the poles where it is very large. Furthermore, the  azimuthal curvature in these regions is also very high. Near the caps deviation from the polar direction is therefore also very costly (as it is much larger than 1, the azimuthal curvature around the equatorial). Thus, a rod  will prefer to avoid regions with high mean curvature (as it is affected by both the polar and azimuthal curvatures). A reasonable criterion is that the curve will remain (up to some penetration depth) in regions satisfying   $\left\|\frac{1}{2}\Tr(S) \right\| \leq1$. By finding the angle $\theta_1$ at which it is equal to $1$  
\begin{align}
1=\left\|\frac{1}{2}\Tr(S) \right\| =\frac{(1+\delta)(2+\delta(2+\delta)\sin^2\theta_1)}{2(1+\delta(2+\delta)\sin^2\theta_1)^{3/2}} \xrightarrow{\delta \gg 1}\frac{1}{2\sin \theta_1}
\end{align} we see that the curve is limited to a region such that $\theta_1 \leq \theta(s)\leq \pi-\theta_1$, primarily pointing along the ellipsoid. Finally, direct calculation shows that $\theta_1 \geq 30^\circ  $
 (monotonically approaching $30^\circ$ in the limit of $\delta \gg 1$). Hence, while in this case $\kappa_g$ is not necessarily small (though it does not need to be too large either), the curve probes much of the surface of the ellipsoid as it traverses about $2/3$ of its length. 

\section{Self avoidance} \label{appndx: self avoidance}
The energy of close packed, almost polar geodesic, path on the ellipsoid, is approximated as the energy of $N=2K+1$ closed paths, so that each path is numbered by $|n| \leq K$, where $n=0$ is the path passing through the poles, and other paths reside in a plane parallel to it. If the filament's thickness is given by $t$ it is not hard to see that the distance of each such filaments from the plane bisecting the ellipsoid through the poles is given by $d= n \cdot t$. The relative distance $y$ is the fraction $\frac{d}{D}$ where $D$ is the semi - minor axis length. For an ordered configuration of maximal widht $d_{max}$ we can show that up to a small correction, $y_{max} \simeq  c$ ,where $c =\frac{L_{tot}  t}{A}$ the packing density, and $L_{tot}$ is the total length of a filament.  As each configuration is planar, it can be written as
\begin{align}
\mathbf{f}(\varphi) = D \left(0,\sqrt{1-y^2}\sin\varphi, \sqrt{1-y^2}(1+\delta)\cos\varphi\right).
\end{align}
A length element is given by 
\begin{align} 
\dt s^2 = D^2\left(1-y^2\right)\left(\cos^2(\varphi)+\left(1+\delta\right)^2 \sin^2(\varphi)\right) \dt \varphi^2,
\end{align}  
and the local curvature of the curve by 
\begin{align}
\kappa= \frac{1+\delta}{D\sqrt{1-y^2}\left(\cos^2(\varphi)+\left(1+\delta\right)^2 \sin^2(\varphi)\right)^{3/2}}.
\end{align} 

Thus the total energy of a single path at $y$ is 
\begin{align} 
E(y)= \frac{1}{\sqrt{1-y^2}} E(\omega=0) 
\end{align}
 where $E(\omega=0)$ is the energy of a polar path. The legnth of a single path is 
 \begin{align} 
 L(y)= \sqrt{1-y^2} L(0)
 \end{align} ($L(0)$ being the length of a polar path). So that 
\begin{align} 
\bar{E}_{SA} = \frac{\displaystyle\int E(y) \dt y}{\displaystyle\int L(y) \dt y} \simeq \frac{2 \arcsin( c)}{ 4\pi c \sqrt{1-  c^2}+\arcsin( c)}  \times \bar{E}(\omega=0).
\end{align}


\bibliography{C:/Users/Doron/Documents/PhD_work/Papers_And_Notes/Bibtex/short,C:/Users/Doron/Documents/PhD_work/Papers_And_Notes/Bibtex/elastic_stat,footnotes}

\end{document}